\documentclass[runningheads]{llncs}
\usepackage[T1]{fontenc}
\usepackage{amsmath}
\usepackage{amssymb}  
\usepackage{pifont}
\usepackage{bm}
\usepackage{booktabs} 
\usepackage{algorithm}
\usepackage{algpseudocode}
\usepackage{multirow}
\usepackage{xcolor}
\usepackage{tabularx}
\usepackage{url}
\usepackage{makecell}
\usepackage{siunitx}
\usepackage{graphicx,verbatim}

\begin{document}

\title{Latent Motion Profiling for Annotation-free Cardiac Phase Detection in Adult and Fetal Echocardiography Videos}

\author{Yingyu Yang\inst{1}\and
Qianye Yang\inst{1}\and
Kangning Cui\inst{1,2}\and Can Peng\inst{1}  \and Elena D'Alberti\inst{3,4}\and Netzahualcoyotl Hernandez-Cruz \inst{1} \and Olga Patey\inst{4} \and Aris T. Papageorghiou\inst{4} \and J. Alison Noble\inst{1}}

\authorrunning{Y. Yang et al.}

\institute{Department of Engineering Science, University of Oxford 
\and Hong Kong Centre for Cerebro-Cardiovascular Health Engineering\and Department of Maternal and Child Health and Urological Sciences, Sapienza University of Rome\and
Nuffield Department of Women’s and Reproductive Health, University of Oxford}
\maketitle

\begin{abstract}
The identification of cardiac phase is an essential step for analysis and diagnosis of cardiac function. Automatic methods, especially data-driven methods for cardiac phase detection, typically require extensive annotations, which is time-consuming and labour-intensive. In this paper, we present an unsupervised framework for end-diastole (ED) and end-systole (ES) detection through self-supervised learning of latent cardiac motion trajectories from 4-chamber-view echocardiography videos. Our method eliminates the need for manual annotations—including ED ES indices, segmentation, or volumetric measurements—by training a reconstruction model to encode interpretable spatiotemporal motion patterns. Evaluated on the EchoNet-Dynamic benchmark, the approach achieves mean absolute error (MAE) of 3 frames (58.3 ms) for ED and 2 frames (38.8 ms) for ES detection, matching state-of-the-art supervised methods. Extended to fetal echocardiography, the model demonstrates robust performance with MAE 1.46 frames (20.7ms) for ED and 1.74 frames (25.3ms) for ES, despite the fact that the fetal heart model is built using non-standardized heart views due to fetal heart positioning variability. 
Our results demonstrate the potential of the proposed latent motion trajectory strategy for cardiac phase detection in adult and fetal echocardiography. This work advances unsupervised cardiac motion analysis, offering a scalable solution for clinical populations lacking annotated data. Code will be released at \url{https://github.com/YingyuYyy/CardiacPhase}.

\footnote{This work has been early-accepted to MICCAI 2025. This version is the submitted manuscript prior to peer review.}
\end{abstract}

\section{Introduction}

Echocardiography is a widely used clinical modality for assessing cardiac function due to its low cost and absence of ionizing radiation. Precise identification of cardiac phases, particularly two important key-points: end-diastole (ED) and end-systole (ES), is critical to derive quantitative functional parameters such as the ejection fraction and ventricular longitudinal strain in adults~\cite{lang2015recommendations}, as well as to perform various measurements for fetal heart assessment~\cite{crispi2013ultrasound}. ED and ES also enable motion pattern alignment for spatiotemporal group analysis~\cite{mcleod2015spatio}.
ED is defined as the frame before mitral valve closure and ES marks aortic valve closure~\cite{mada2015define}. 
In the 4-chamber (4CH) view , one of the most informative echocardiographic planes, the current clinical practice relies on visual assessment by sonographers or cardiologists to identify these key frames, a subjective process that introduces inter-observer variability and depends heavily on operator experience.
  
Automated cardiac phase detection in 4CH echocardiography has employed direct and indirect approaches. Direct data-driven methods use deep learning architectures to predict ED/ES end-to-end. Spatio-temporal information is processed through hybrid CNN-RNN frameworks \cite{dezaki2018cardiac,lee2020automatic}, Vision Transformers~\cite{reynaud2021ultrasound} to estimate temporal ED/ES probability. Authors of~\cite{pu2024hfsccd} apply a multi-task CNN network to perform fetal heart localization, phase detection and standard plane classification. Indirect methods derive ED/ES from intermediate prediction such as left ventricle (LV) volume curves~\cite{li2023semi} or temporal segmentation masks~\cite{zeng2023maef}, extracting phases from extrema. Both strategies require extensive manual annotations (e.g., ED/ES indices, LV segmentation), limiting their applicability in annotation-scarce situations.

Unsupervised ED/ES detection methods offer promising solutions to reduce annotation dependency. Early approaches like those by Gifani and Shalbaf~\cite{gifani2010automatic,shalbaf2015echocardiography} mapped apical 4CH echocardiography videos to low-dimensional manifolds via locally linear embedding (LLE), detecting phases through density analysis. However, their case-specific embeddings lack interpretability and generalizability. Laumer et al.~\cite{laumer2020deepheartbeat} advanced this by learning a circular latent trajectory from cardiac videos, with ED/ES phases semantically aligned to specific arc regions in the learned latent circle. While innovative, this method cannot pinpoint exact ED/ES timings, limiting potential clinical utility. More recently, a training-free unsupervised approach based on left ventricular expansion-contraction dynamics, called DDSB, was introduced~\cite{ddsb}. Though computationally efficient, it struggles with videos containing short ED/ES intervals, hindering broad applicability.

We propose LMP (Latent Motion Profiling), an annotation-free framework for unsupervised cardiac phase detection in apical 4CH echocardiography. Our key contributions are as follows.

\begin{itemize}
\item \textbf{Unsupervised Interpretability:} A novel latent motion subspace where cardiac dynamics are encoded into two orthogonal physiologically meaningful directions (septal and lateral movement). ED and ES occupy distinct trajectory regions, enabling annotation-free detection.
\item \textbf{Label-Free Generalization:} By training solely on video reconstruction, LMP eliminates dependence on ED/ES indices, segmentations, or clinical measurements, overcoming a critical bottleneck in fetal and paediatric applications where annotations are scarce. 
\item \textbf{Cross-Population Robustness:} Consistent performance in adult and fetal echocardiography, achieving: supervised-level accuracy on adult data benchmarks (MAE: 3.0 ED / 2.0 ES frames), and on fetal results (1.46 ED / 1.74 ES frames). This demonstrates the first unified framework for cardiac phase detection in both populations.  
\end{itemize} 
 
\section{Method}
\subsection{Echocardiography Motion Profiling via Video Reconstruction
}

\begin{figure}[t]
\includegraphics[width=\textwidth]{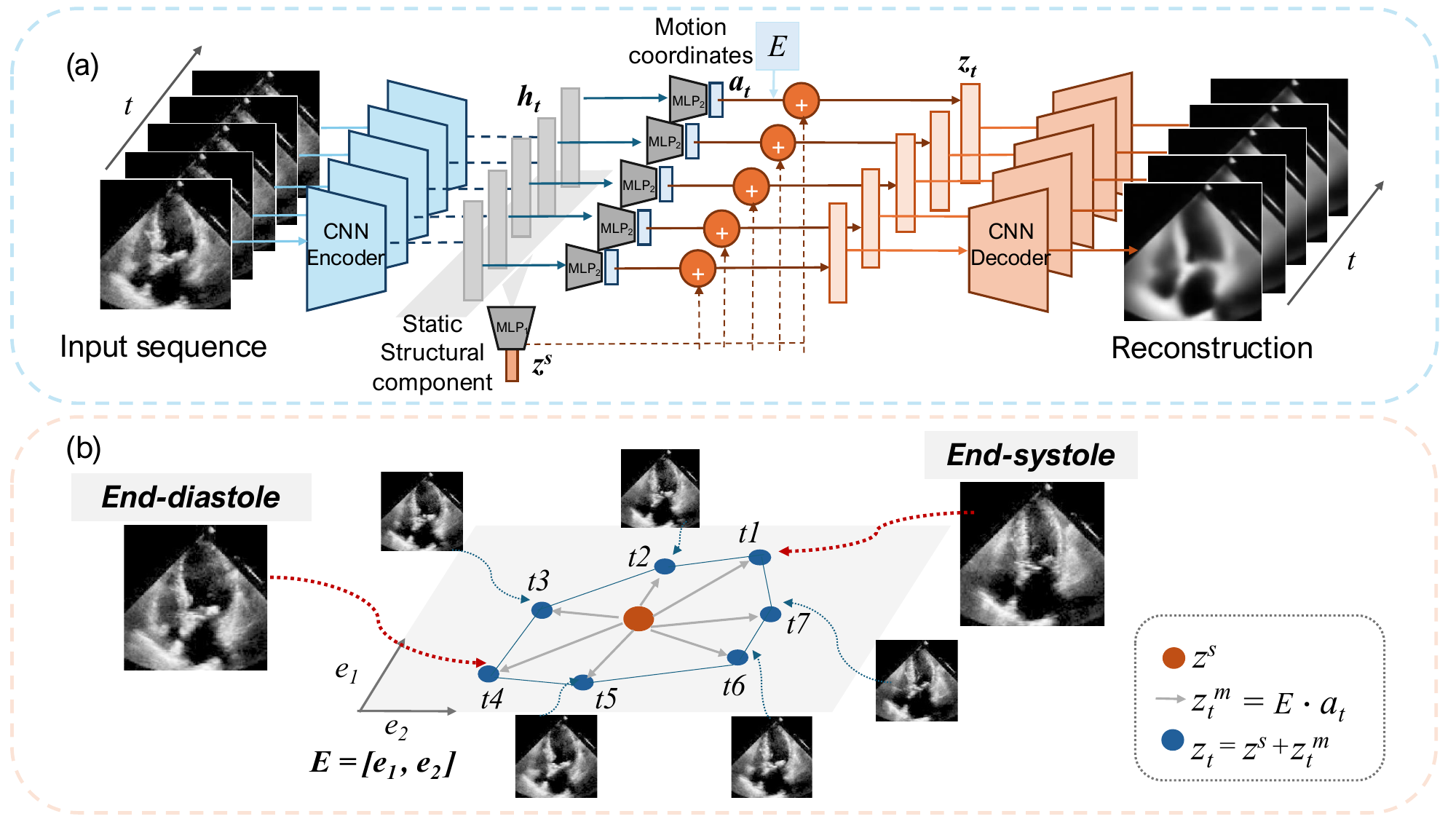}
\caption{\textbf{Latent Cardiac Motion Profiling via Frame-wise Reconstruction.} 
    \textbf{(a)} Structure-motion decomposition: Input frame $x_t$ is encoded as $z_t = z^s + \mathbf{E}\bm{a}_t$, where $z^s \in \mathbb{R}^D$ represents static anatomy and $\mathbf{E}\bm{a}_t$ models motion in orthogonal subspace $\mathbf{E} = [\bm{e}_1, \dots, \bm{e}_K]$.  
\textbf{(b)} Motion trajectory: Temporal evolution of $\{\bm{a}_t\}$ in $\mathbb{R}^K$ ($K = 2$ in our case) reveals ED/ES phases as geometric landmarks, enabling unsupervised detection through trajectory analysis.
    } \label{fig_method}
\end{figure}
Given a video sequence \( X \in \mathbf{R}^{T \times H \times W} \), where \( T \), \( H \), and \( W \) denote the temporal, spatial height, and spatial width dimensions, respectively. Our objective is to learn a temporally coherent sequence of low-dimensional latent vectors \( \{ z_t \}_{t=1}^T \in \mathbf{R}^D \) for video reconstruction through an autoencoder framework. Inspired by the low-dimensional representation analysis of cardiac deformation\cite{rohe2018low,ryser2022anomaly}, we propose a \textit{structure-motion decomposition} approach. Specifically, for each frame \( x_t \), the latent representation \( z_t \) is decomposed into two components:
\begin{equation} \label{eq:structure_motion}
z_t = z^s + z^m_t 
\end{equation}
\noindent\textbf{Static structural component}: $z^s \in \mathbf{R}^D$ encodes subject-specific anatomical features and accounts for structural variations between distinct videos. It is computed as the mapping of the mean latent embedding through a MLP $z^s = \textbf{MLP}_1 (\frac{1}{T}\sum_{t=1}^T h_t)$, where $h_t$ is the frame-wise embedding via a convolutional encoder (Fig. \ref{fig_method}(a)).
    
\noindent\textbf{Dynamic motion component}: \( z^m_t \in \mathbf{R}^D \) captures temporally evolving cardiac motion relative to \( z^s \) (Fig. \ref{fig_method}(b)). \( z^m_t \) lies in a low-rank subspace spanned by an orthogonal basis \( \mathbf{E} = [\bm{e}_1, \bm{e}_2, \dots, \bm{e}_K] \in \mathbf{R}^{D \times K} \), where \( K \ll D \). The basis vectors \( \{ \bm{e}_k \}_{k=1}^K \), optimized during training, define a motion subspace that preserves trajectory smoothness and physiological plausibility. Temporal dynamics within this subspace are governed by learnable coefficients \( \bm{a}_t \in \mathbf{R}^K \) through another MLP : $\bm{a}_t = \textbf{MLP}_2 (h_t)$~\cite{wang2022latent}. The motion component is computed $z_t^m = \mathbf{E} \bm{a}_t$.

Finally, the composite latent $z_t$ is decoded by the decoder $\mathcal{D}$ to reconstruct the input frame.
To enforce consistency between the decomposed latent space and pixel-level video dynamics, we apply the following loss function for optimization: 
\begin{equation}
    \mathcal{L} (X) = \frac{1}{T} \sum_{t=1}^T || \mathcal{D} (z^s) - x_t||_2^2 +  \sum_{t=1}^T||\mathcal{D}(z_t) - x_t||^2_2 
\end{equation}
\noindent where the first term aims to find the $z^s$ that reconstructs the Fréchet mean image of a given sequence and the second term focuses on the reconstruction of each temporal frame.

\begin{figure}[t]
\includegraphics[width=\textwidth]{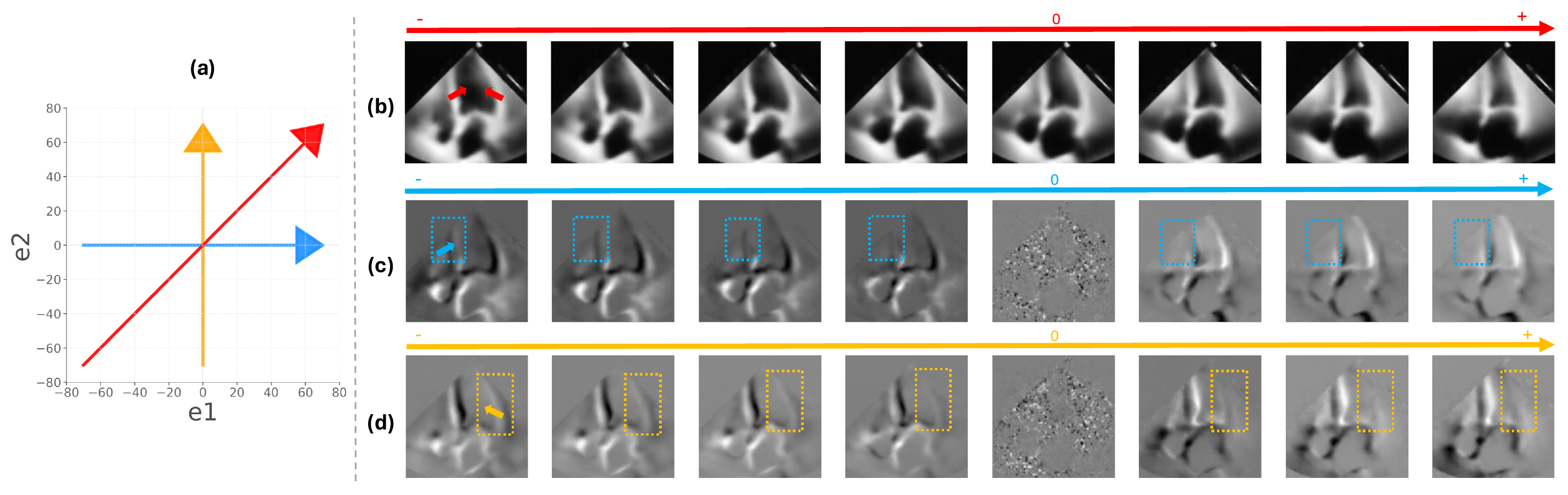}
\caption{\textbf{Latent motion disentanglement visualization in the motion subspace spanned by $\mathbf{E} = [\bm{e}_1, \bm{e}_2]$.}
(\textbf{a}) Three latent motion trajectories: two axes-specific motions (orange and blue lines) and their combination (red line). (\textbf{b}) Reconstruction of the red trajectory showing cardiac contraction with left ventricle volume decreasing and the movement of the septum and lateral wall (red arrows). (\textbf{c}) Reconstruction difference between red and blue trajectories, with minimal differences in the dashed region indicating e1 axis correlation with septal movement (blue arrow). (\textbf{d}) Reconstruction difference between red and orange trajectories, with minimal differences in the dashed region indicating e2 axis correlation with lateral movement (orange arrow).} \label{fig_disentanglement}
\end{figure}

\begin{algorithm}[t]
\caption{ED/ES detection from latent motion trajectory}
\label{alg:pca_extremes}

\textbf{Input:} Trajectory point coordinates $\{\bm{a}_t = (a^1_t, a^2_t)\}_{t=1}^T$ \\
\textbf{Output:} End-diastole indices $\text{Group}_{ED}$, End-systole indices $\text{Group}_{ES}$

\begin{algorithmic}[1]
\State \textbf{Principal orientation detection:}
\begin{itemize}
    \item Compute normalized displacement vectors: $d(t) = \frac{\bm{a}_{t+1} - \bm{a}_t}{||\bm{a}_{t+1} - \bm{a}_t||}$
    \item Apply RANSAC~\cite{fischler1981random} to determine inlier displacement vectors and compute the main orientation axis $\mathbf{v} = (v_1, v_2)$ using PCA 
\end{itemize}

\State \textbf{Trajectory projection to the principal orientation axis:}
    \begin{itemize}
        \item Normalize the main orientation direction $\bar{\mathbf{v}} = \frac{\mathbf{v}}{||\mathbf{v}||}$ and compute the mean position of the trajectory $\mu$
        \item Projected trajectory:  $\mathbf{s} =\{s_t\}_{t=1}^T, s_t = (a_t - \mu) \cdot \bar{\mathbf{v}}$
    \end{itemize}

\State \textbf{Projected trajectory preprocessing:}
    \begin{itemize}
        \item Smoothing: Apply Savitzky-Golay filter to $\mathbf{s}$ with window $w=8$, order $p=2$
        \item Remove baseline wander: Apply high-pass filter (cutoff 0.5 Hz) if low-frequency power ratio exceeds 0.1
        \item Get filtered signal $\mathbf{s}_{filt}$
    \end{itemize}

\State \textbf{ED/ES indices detection:}
    \begin{itemize}
        \item Find peaks and valleys in $\mathbf{s}_{filt}$ using prominence threshold $0.3 (\mathbf{s}_{filt}^{max} - \mathbf{s}_{filt}^{min})$
        \item Assign indices: $\text{Group}_{ES} = \text{Peaks}, \text{Group}_{ED} = \text{Valleys}$
    \end{itemize}
\end{algorithmic}
\end{algorithm}

\subsection{Cardiac Phase Identification in Learned Motion Space}
\begin{figure}[t]
\includegraphics[width=\textwidth]{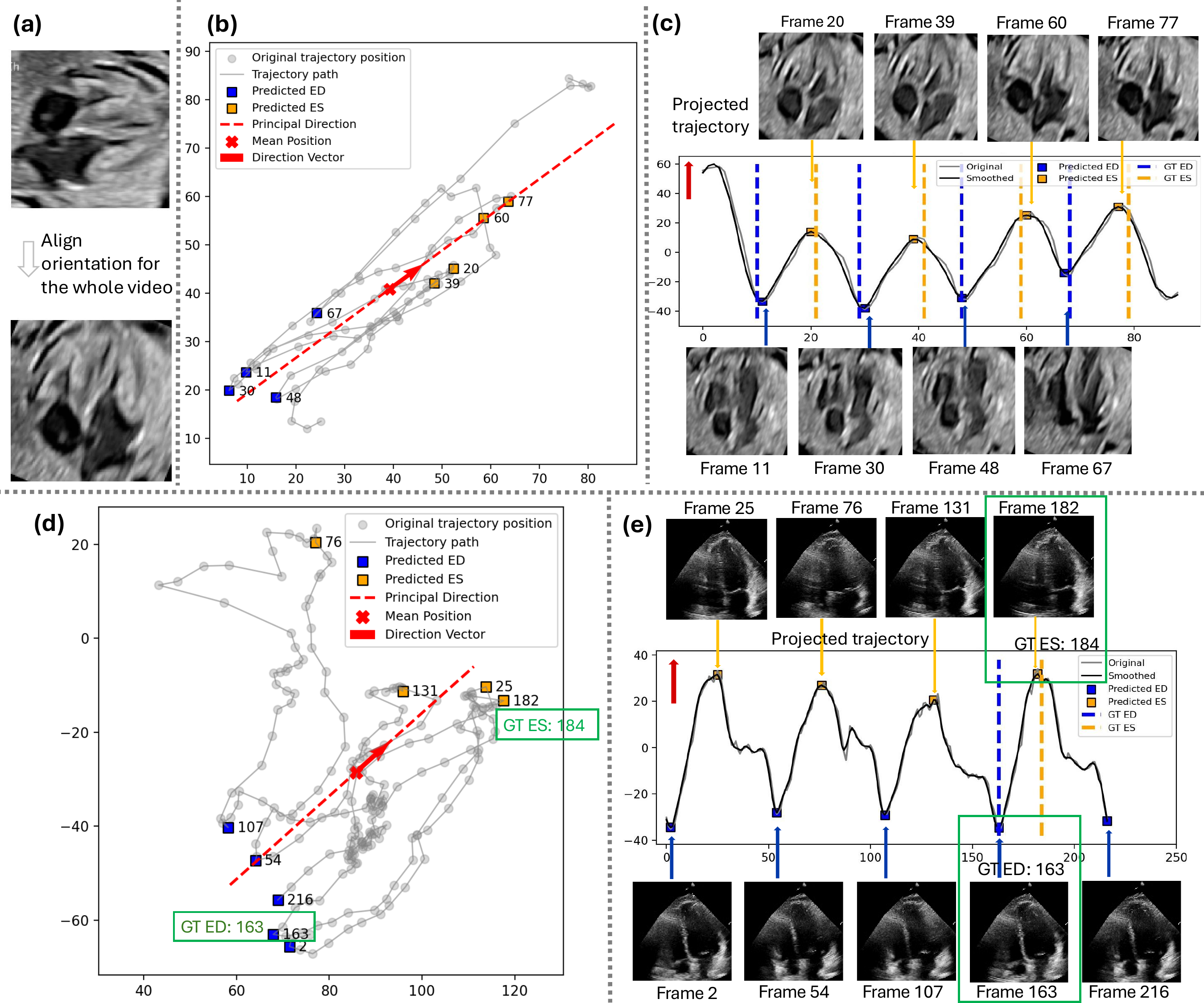}
\caption{\textbf{Latent motion trajectory for ED/ES indices detection.}
(\textbf{a-c}) A fetal test example. (a) Align fetal 4CH view to canonical apical orientation. (b) 2D latent motion trajectory. (c) Following Algorithm \ref{alg:pca_extremes}, the motion trajectory is projected to the principal direction. Peaks of the projected trajectory indicate ES and valleys indicate ED. Predicted ED/ES image frames are shown. (\textbf{d-e}) An adult test example. 
} 
\label{fig_trajectory}
\end{figure}

Cardiac motion during one cardiac cycle, as captured by the 4CH view, demonstrates the contraction of the ventricle (with the valve moving toward the apex and the ventricular volume decreasing) and the relaxation of the ventricle (with the valve moving away from the apex and the ventricular volume increasing). In this work, we choose to set the motion subspace with \( K = 2 \) due to its simplicity and the interpretability of the disentanglement of cardiac motion in the 4CH view. We find that through the reconstruction task, the latent motion subspace learns, in an unsupervised manner, the movement of the septal and lateral heart wall, respectively (Fig. \ref{fig_disentanglement}). This results in a latent motion trajectory following a back-and-forth motion pattern, with the extremities of the trace representing the points of motion phase change, which are closely related to ED and ES (Fig. \ref{fig_trajectory}(b) and Fig. \ref{fig_trajectory}(d)). We apply Algorithm~\ref{alg:pca_extremes} to detect ED and ES indices of a given motion trajectory.

\section{Experiments}
We conducted unsupervised training experiments on two independent echocardiography video datasets: (1) a public adult dataset and (2) a private fetal dataset. Both experiments shared core architectures but employed distinct preprocessing and evaluation protocols to address domain-specific challenges.

\subsection{Adult Echocardiography Experiment}  
The adult analysis used the EchoNet-Dynamic dataset~\cite{ouyang2020video}, containing 10,030 apical 4CH view videos. They have annotated one single ED and ES frames for each video. We followed the train/validation/test splits in~\cite{ouyang2020video}.  

\noindent \textbf{Evaluation} 
Since only one ground truth ED/ES was provided, we calculated the mean absolute error (MAE) between the ground truth and the temporally closest prediction.

\subsection{Fetal Echocardiography Experiment}  
Our private fetal dataset comprised 449 four-chamber view B-mode videos of normal fetal hearts from 313 healthy participants imaged during the second trimester of gestation. We manually cropped the heart area and aligned it with apical 4CH view orientation (Fig. ~\ref{fig_trajectory}. (a)). A fetal cardiologist annotated 44 test videos generally containing 4 cardiac cycles, providing continuous ED/ES labels. The remaining 405 videos were split 85:15 for training and validation.

\noindent \textbf{Training strategies} Two training approaches were implemented: 1) \textit{Standard Training}: Model initialization as for the adult echocardiography experiment. 
2) \textit{Transfer Learning}: Fine-tuning the best-performing adult echocardiography model on fetal data, preserving learned motion priors. 

\noindent \textbf{Evaluation}
We adopted three complementary metrics:(1) \textit{GT-centric MAE} focuses on temporal alignment completeness by measuring the error between each ground truth (GT) frame and its closest prediction. It penalizes failures to detect true events. (2) \textit{Prediction-centric MAE} evaluates detection precision by computing the error between each prediction and its nearest GT annotation. It penalizes extraneous or false-positive detections. (3) \textit{Matched-pair MAE} assesses localization accuracy for temporally aligned pairs: GT and prediction pairs were matched if their temporal offset was less than 50\% of the mean cardiac cycle length (derived from test data). This metric isolates errors only for confident matches, avoiding noise from mismatches.

\subsection{Implementation Details}  
We preprocessed videos by resizing frames to $128\times128$ pixels and normalizing intensities to [0,1]. We randomly extracted 50-frame clips from each video for training efficiency and temporally downsampled them to 25 frames. The reconstruction model was trained for 500 epochs using the Adam optimizer (learning rate = 0.001 for adult data and 0.0001 for fetal data, batch size = 32) with extensive data augmentation: random brightness/contrast adjustments ($\pm20\%$), Gaussian blurring ($\sigma\in[0.25,1.5]$), additive noise ($\sigma=0.01$), rotation ($\pm60^\circ$), translation (10\% offset range), scaling (0.85--1.15), and horizontal flipping. Model selection was guided by validation split performance. Inference used full-length original videos without temporal sampling. 

\section{Results}
\subsection{Cardiac Phase Detection in Adult Echocardiography}

Table \ref{tab_echonet_result} summarizes cardiac phase detection results for the EchoNet-Dynamic test split, comparing our method with existing approaches from~\cite{li2023semi,zeng2023maef,reynaud2021ultrasound}. Notably, MAEF and UVT did not report MAE in milliseconds. To enable cross-method comparison, we converted frame-based MAE to temporal error (ms) using a dataset's mean frame rate of 51.52 fps (19.42 ms per frame), as 80\% of test samples were recorded at 50 fps. For DDSB~\cite{ddsb}, which originally reported results on a subset of EchoNet-Dynamic, we re-evaluated their official implementation on the test split to ensure fair comparison - this accounts for discrepancies from their published results. Our unsupervised method achieves MAEs of 3 frames (58.3 ms) for end-diastole (ED) and 2 frames (39.8 ms) for end-systole (ES) detection. This performance not only surpasses the unsupervised approach, but also matches the accuracy of state-of-the-art supervised methods. An example of detection shown in Fig. \ref{fig_trajectory}(e). 

\begin{table}[t]
\begin{minipage}{\textwidth}
  \centering
  \caption{Cardiac phase detection in adult echocardiography  (1276 test videos of~\cite{ouyang2020video}).
  }
  \label{tab_echonet_result}
  \resizebox{\textwidth}{!}{\begin{tabular}{rcccccc}
    \toprule
    \multirow{2}{*}{\textbf{Method}} & 
    \multirow{2}{*}{\textbf{Supervision}} & 
    \multirow{2}{*}{\textbf{Annotation}} &
    \multicolumn{2}{c}{\textbf{MAE (frames)}} & 
    \multicolumn{2}{c}{\textbf{MAE (ms)}} \\
    \cmidrule(lr){4-5} \cmidrule(lr){6-7}
    & & & \textbf{ED} & \textbf{ES} & \textbf{ED} & \textbf{ES} \\
    \midrule
    R3D-50~\cite{li2023semi} & Yes  & LV volume & $2.1(2.5)$ & $1.7(2.7)$ & $40.2(48.9)$ & $32.6(51.6)$ \\
    MAEF~\cite{zeng2023maef} & Yes & LV segmentation & $2.3(2.3)$ & $2.4(2.2)$ & $44.5(44.2)$ & $46.1(42.3)$ \\
    UVT~\cite{reynaud2021ultrasound} & Yes & EF,ED,ES & $7.2(12.9)$ & $3.4(6.8)$ & $139.2(250.4)$ & $65.0(131.9)$ \\
    \midrule
    DDSB~\cite{ddsb}\footnote{\scriptsize{Rerun using official implementation} 
    } & \textbf{No}& \textbf{None} & $4.3(4.8)$ & $8.9(9.5)$ & $83.4(93.3)$ & $175.4(185.4)$\\
    Ours & \textbf{No}& \textbf{None} & $3.0(3.3)$ & $2.0(2.2)$ & $58.3(66.4)$ & $39.8(43.0)$ \\ 
    \bottomrule
  \end{tabular}}
\end{minipage}
\end{table}

\subsection{Cardiac Phase Detection in Fetal Echocardiography}

\begin{table}[t]
  \centering
  \caption{Cardiac phase detection in fetal echocardiography (44 test videos)}
  \label{tab_fetal_result}
  \resizebox{\textwidth}{!}{\begin{tabular}{rccccccc}
    \toprule
    \multirow{2}{*}{\textbf{Training}} & 
    \multirow{2}{*}{\textbf{Metric}} & 
    \multicolumn{2}{c}{\textbf{\#Samples}} &
    \multicolumn{2}{c}{\textbf{MAE (frames)}} & 
    \multicolumn{2}{c}{\textbf{MAE (ms)}} \\
    \cmidrule(lr){3-4} \cmidrule(lr){5-6} \cmidrule(lr){7-8}
    & & \textbf{ED} & \textbf{ES} & \textbf{ED} & \textbf{ES} & \textbf{ED} & \textbf{ES} \\
    \midrule
    \multirow{3}{*}{Standard} & GT-centric & 165 & 161 & $3.22(7.17)$ & $3.78(9.45)$ & $46.3(103.0)$ & $53.7(126.3)$\\
    & Prediction-centric &157 & 154 & $2.16(4.68)$ & $2.38(4.06)$ & $31.3(69.6)$ & $34.5(57.8)$  \\
    & Match-pair & 155 & 152 & $1.46(1.38)$ & $1.77(1.36)$ & $21.0(19.7)$ & $25.7(19.4)$\\
    \midrule
    \multirow{3}{*}{Transfer} & GT-centric & 165 & 161&$3.21(8.85)$ & $2.68(5.20)$ & $48.5(142.6)$ & $39.6(77.5)$\\ 
    & Prediction-centric &159 & 157& $2.17(4.81)$ & $2.20(3.79)$ & $31.3(71.9)$ & $31.8(52.8)$\\
    
    & Match-pair & 157 & 155 & $1.46(1.26)$ & $1.74(1.48)$ & $20.7(17.9)$ & $25.3(21.7)$\\
    \bottomrule
  \end{tabular}}
\end{table}

Table \ref{tab_fetal_result} presents fetal cardiac phase detection results using our unsupervised method. Transfer learning from adult echocardiography preserved learned ED/ES motion patterns, achieving a reduction of 1.1 frames in ES MAE compared to standard fetal training for GT-centric metrics. Both approaches learned interpretable cardiac motion trajectories without any annotation from real-world fetal data. 
Our method has correctly detected 95\% ED frames and 96\% ES frames. Match-pair MAE (ED: 1.46 frames, ES: 1.74 frames) shows strong alignment with the ground truth. An example of detection shown in Fig. \ref{fig_trajectory}(c).

\section{Conclusion and Discussion}
In this work, we propose an unsupervised cardiac phase detection framework that learns interpretable latent motion trajectories from echocardiography videos without requiring annotations (ED/ES indices, LV segmentation, or volume measurements). Validated for both adult (apical 4CH views) and fetal echocardiography, our approach achieves performance comparable to supervised methods. Notably, this work establishes a foundation for annotation-free cardiac motion analysis for diverse clinical populations. Finally, in this work, we mainly focus on apical 4CH view videos (where we have aligned the fetal 4CH view videos to apical orientation), while in reality, positional variability often results in non-apical 4CH views (e.g., basal/transverse) in the fetal cardiac examination, requiring future work to address view-invariant motion modelling. Beyond phase detection, the learned latent space may enable novel applications, including motion artifact disentanglement and pathology-sensitive trajectory clustering, which will be investigated in future work. 

\begin{credits}
\subsubsection{\ackname} This work was partly supported by the InnoHK-funded Hong Kong Centre for Cerebro-cardiovascular Health Engineering (COCHE) Project 2.1 (Cardiovascular risks in early life and fetal echocardiography). Co-authors J. Alison Noble and Aris Papageorghiou were supported by the Oxford Partnership Comprehensive Biomedical Research Centre with funding from the NIHR Biomedical Research Centre (BRC) funding scheme.

\end{credits}

\bibliographystyle{splncs04}
\bibliography{ref}

\end{document}